\documentstyle[epsfig]{aipproc}
\def\kms{km$\,{\rm s}^{-1}$}

\begin{document}
\title{Far Ultraviolet Spectroscopy of a Nonradiative Shock in
the Cygnus Loop: Implications for the Postshock Electron-Ion
Equilibration}

\author{Parviz Ghavamian$^1$, John C. Raymond$^2$ and William P. Blair$^3$}
\address{$^1$ Department of Physics and Astronomy, Rutgers University,
Piscataway NJ, 08854\\
$^2$ Harvard-Smithsonian Center for Astrophysics, Cambridge, MA 02138\\
$^3$ Department of Physics and Astronomy, Johns Hopkins University,
Baltimore, MD 21218\\}

\maketitle

\begin{abstract}
We present far ultraviolet spectra of a fast ($\sim\,$300 \kms) nonradiative
shock in the Cygnus Loop supernova remnant.  Our observations were performed with
the Far Ultraviolet Spectroscopic Explorer (FUSE) satellite, covering the
wavelength range 905$-$1187 \AA.  The purpose of these observations
was to estimate the degree of electron-ion equilibration by (1) examining the
variation of O~VI $\lambda\lambda$1032, 1038  intensity with position behind the shock, and (2)
measuring the widths of the OVI lines.  We find significant absorption near the center
of the O~VI $\lambda$1032 line, with less absorption in O~VI $\lambda$1038.
The absorption equivalent widths imply an O~VI column density greatly exceeding that of interstellar
O~VI along the Cygnus Loop line of sight.  We suggest that the absorption may be due to
resonant scattering by O~VI ions within the shock itself.  The widths of the
O~VI emission lines imply efficient ion-ion equilibration, in agreement with predictions from
Balmer-dominated spectra of this shock.
\end{abstract}

\section*{Introduction}

A shock wave is termed nonradiative if the cooling time of the postshock
gas exceeds the dynamical time scale.  Nonradiative shocks lack cooling zones, therefore
the optical and ultraviolet line emission is excited close to the shock front.
The emission line fluxes and line widths are sensitive to the shock velocity and the degree of postshock
electron-ion/ion-ion equilibration.  This property makes the optical and ultraviolet lines valuable tools
for investigating collisionless heating in high
Mach number, low density shock waves.

For an ion of mass $m_{i}$, the jump conditions for a strong shock in an ideal gas
medium ($\gamma\,=\,\frac{5}{3}$) give a postshock temperature
\begin{equation}
T_{i}\,=\,\frac{3}{16}\,\frac{m_{i}\,V_{s}^{2}}{k}
\end{equation}
In an unequilibrated shock, the electron and proton temperatures are in the
ratio 1/2000:1, while the temperatures of O ions and protons are in
the ratio of 16:1.  In contrast, the temperatures of all three species are equal
in an equilibrated shock.  Collisionless heating can produce equilibrations
anywhere between these two extremes.  In principle, the amount of collisionless heating
from plasma waves, MHD turbulence and other processes is sensitive to such parameters as
the magnetosonic Mach number and angle between the magnetic field and shock front.  Understanding
this relationship is of critical importance in the interpretation of X-ray, ultraviolet and optical
data of high Mach number collisionless shocks.

A nonradiative shock in partially neutral gas emits
a pure Balmer line spectrum generated by collisional excitation.  Each
Balmer line consists of a narrow component produced by excitation of cold 
H~I and a broad component produced by proton-H~I charge exchange.  The shock
velocity and electron-ion equilibration can be gauged from the broad component
line width and ratio of broad component to narrow component flux \cite{[1]}\cite{[2]}.
The ion-ion equilibration can be
gauged in turn by measuring the width of the O~VI $\lambda\lambda$1032, 1038 lines
and the variation of O~VI flux with position behind the shock. 

\section*{Observations}

The target of our FUSE observations was Cygnus P7, a nonradiative shock in the NE
corner of the Cygnus Loop supernova remnant\cite{[1]}.  We acquired ultraviolet spectra of Cygnus P7
in June 2000 for a total of 64 ks (all during orbital night) and four
slit positions behind the shock (Fig. 1).
\begin{figure} 
\centerline{\epsfig{file=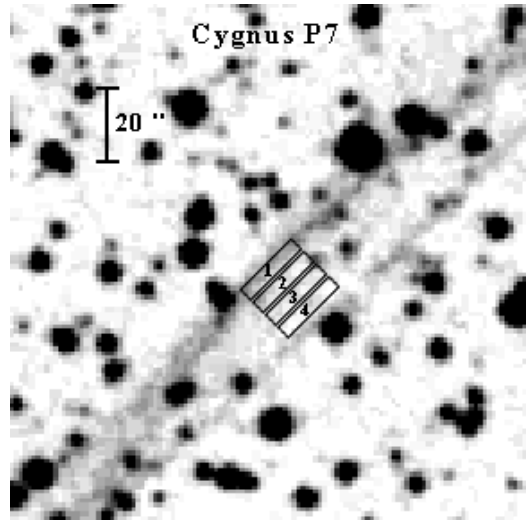,height=2.7in,width=2.7in}}
\caption{POSS image of the Cygnus P7 filament.  The field of view is approximately
2.5$^{\arcmin}\times$2.5$^{\arcmin}$ square. The FUSE slit positions are marked (PA\,=\,315$^{\circ}$).}
\label{fig1}
\end{figure}
The spectra presented here were acquired through the 4$^{\arcsec}\,\times\,$20$^{\arcsec}$
medium resolution (MDRS) slit.   Slit position 1 was centered on the crisp Balmer-dominated filament and
placed parallel to it (Fig. 1).  To observe the intensity variation of O~VI, we centered slits
2, 3 and 4 at 5$^{\arcsec}$, 10$^{\arcsec}$ and 15$^{\arcsec}$ behind
the shock, respectively.
The one-dimensional spectra of Positions 1, 2, 3 and 4 appear in Fig. 2.
To minimize the background contribution, these
spectra were extracted from the LiF 1A channel (one of four available with FUSE),
the detector yielding the highest effective area at $\lambda\,>\,$1020 \AA.  The spectral
resolution is 0.08 \AA (25 \kms), corresponding to filled slit emission.
\begin{figure}
\centerline{\epsfig{file=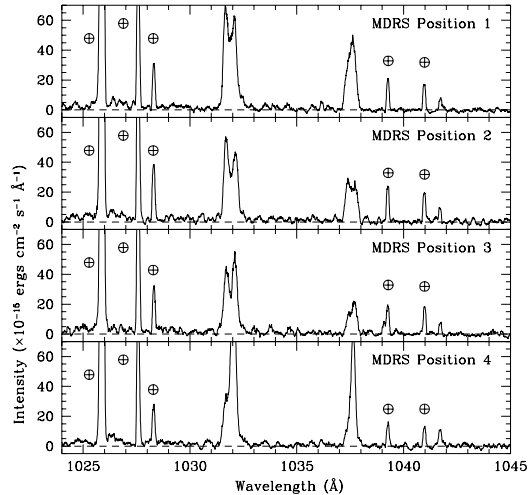,height=2.8in,width=2.8in}}
\caption{One-dimensional FUSE spectra of Cygnus P7 at slit positions 1, 2, 3 and 4.  Spectra
have been smoothed over 0.1 \AA, and are presented without reddening correction.
}
\label{fig2}
\end{figure}

\section*{Analysis}

The 1-D spectra of Cygnus P7 (Fig. 2) reveal three important features:
(1) The O~VI lines are considerably broader than the airglow
lines, confirming the detection of O~VI emission in the Cygnus P7 shock.  The
velocity width of the O~VI $\lambda$1032 line is
$\sim\,$120 \kms.
 Combining this result with the shock velocity of 300$-$365 \kms
predicted from Balmer line spectroscopy of Cygnus P7, we obtain
$T_{e}/T_{O}\,\sim\,$0.7.
 This is roughly consistent with the
electron-proton equilibration $T_{e}/T_{p}\,\sim\,$0.7$-$1.0 predicted by the Balmer
line analysis; 
In an earlier analysis\cite{[1]} of Balmer line emission from Cygnus P7, we measured
an H$\alpha$ broad component width of 262 \kms. This corresponds to a shock velocity of
265$-$365 \kms\, for cases of no equilibration and full
equilibration, respectively. 
(2) the O~VI emission lines are strongly affected by absorption.
The absorption features tend to shift blueward at slit
positions farther and farther behind the shock.  The variation of O~VI
flux with position is masked by the O~VI absorption features; 
(3) the shape of the $\lambda$1038 absorption profile differs from
that of the $\lambda$1032 profile.  This is mainly due to galactic
H$_{2}$ absorption at the position of the O~VI $\lambda$1038 line.

\section*{Interpretation}

There are two possible explanations for the oberved O~VI absorption: either
the bulk of the absorption is produced by interstellar matter (O~VI, H$_{2}$, C~II$^{*}$, etc.)
along the line of sight, or the absorption is produced locally by O~VI ions within the 
shock.  There are three lines of evidence in favor of the latter interpretation:

(1) The velocity width of the O~VI $\lambda$1032 absorption feature is $\sim\,$90 \kms, comparable
to the width of the $\lambda$1032 emission line.  The O~VI column density and temperature required to produce
the observed absorption are N$_{O~VI}\,\gtrsim\,$2$\times$10$^{14}$
cm$^{-2}$ and T$_{O~VI}\,\sim\,$3$\times$10$^{6}$K.  These values are
$\gtrsim\,$10 times larger than the O~VI column density
and temperature expected for the ISM along the Cygnus Loop line of sight\cite{[3]}.
(2) The progressively larger blueshift of the O~VI absorption lines with distance
behind the shock cannot be explained by interstellar O~VI absorption.
(3) The line center ratio I$_{1038}$(v=0)/I$_{1032}$(v=0) varies significantly with
slit position, from around 0.6 (Position 2) to 1.1 (Position 4).  These localized
variations are consistent with O~VI resonant scattering within the shock.

The appearance of the O~VI spectrum may be due to a combination of internal resonant scattering and
a curved shock geometry.  On the one hand, the O~VI column density and temperature obtained from the FUSE spectra
are consistent with the range of shock velocities (300$-$365 \kms\, from the Balmer lines) and preshock densities
(n$\,\sim\,$0.1$-$0.3 cm$^{-3}$ from ROSAT X-ray observations) estimated for Cygnus P7.
On the other hand, a curved shock geometry could explain both the increasing value of
I$_{1038}$(v=0)/I$_{1032}$(v=0) and the increasing absorption blue shift with position behind
the shock.  In a curved shock, O~VI photons produced farther downstream
could be absorbed by O~VI ions on the near (blue shifted) side of the shock.

\section*{Conclusions}

In FUSE observations of a nonradiative shock in the NE Cygnus Loop, we find
O~VI emission line widths consistent with moderate to high electron-ion equilibration.  We
also find strong O~VI absorption near the centers of the O~VI emission lines.
A quantitative explanation of the interplay between the O~VI emission and resonant scattering requires
(1) a careful modeling of the shock geometry, and (2) a Monte Carlo simulation to
adequately treat resonant scattering at moderate optical depths (0\,$\leq\,\tau\,
\leq\,$1).  We will address these issues in future work.

This research project is supported by NASA grant NAG5-9019.


\end{document}